\documentclass{article}

\usepackage{PRIMEarxiv}

\usepackage[utf8]{inputenc} 
\usepackage[T1]{fontenc}    
\usepackage{hyperref}       
\usepackage{url}            
\usepackage{booktabs}       
\usepackage{amsfonts}       
\usepackage{nicefrac}       
\usepackage{microtype}      
\usepackage{lipsum}
\usepackage{graphicx}
\graphicspath{{media/}}     
\usepackage{amsmath}
\usepackage{amssymb}
\usepackage{cases}
  
\title{A data-driven approach to modeling brain activity using differential equations}

\author{
  Kuratov A. \\
  Faculty of Computer Science \\
  HSE University  \\
  Moscow\\
  \texttt{kuratov.andrew@gmail.com} \\
}

\begin{document}
\maketitle

\begin{abstract}
This research focuses on an innovative task of extracting equations from incomplete data, moving away from traditional methods used for complete solutions. The study addresses the challenge of extracting equations from data, particularly in the study of brain activity using electrophysiological data, which is often limited by insufficient information. The study provides a brief review of existing open-source equation derivation approaches in the context of modeling brain activity. The section below introduces a novel algorithm that employs incomplete data and prior domain knowledge to recover differential equations. The algorithm's practicality in real-world scenarios is demonstrated through its application on both synthetic and real datasets.
\end{abstract}


\section{Introduction}

One area of focus in machine learning development is the recovery or extraction of differential equations that govern the observed data. This process bridges the gap between fundamental and applied scientific disciplines. The availability of extensive data describing complex models in fields such as biology, medicine, physics, economics, and sociology facilitates the improvement of existing models, the development of new ones, and their subsequent verification. This is due to the advancement of methods for recovering governing equations.

Traditionally, methods have focused on extracting equations from fully resolved data using a variety of libraries and programs for recovering ordinary and partial differential equations. Section \ref{sec:RelWorks} provides a classification of these tools, highlighting notable open-source libraries and their applications in equation recovery.

However, this rapidly evolving landscape presents challenges when data is incomplete or when the underlying model exceeds the scope of observed variables. This paper aims to address this issue, specifically in the context of brain modelling, with a focus on reconstructing alpha-wave generation models from limited data. By tackling the problem locally, the research seeks to bridge the gap between incomplete data and the desired equation representation.

\section{Related works}\label{sec:RelWorks}
In methods for extracting governing equations from data, we can distinguish two
large classes. The first class includes algorithms based on sparse regression and the second class includes algorithms based on neural networks. There are also other methods.
\subsection{Methods based on sparse regression}
Algorithms in this class use the sparse regression method to recover various types of differential equations. However, the dependencies in these equations are strictly taken from a priori given sets of functions. These sets can be large, which speeds up calculations. The algorithm requires noise filtering and noise- and emission-resistant methods for deriving derivatives from the sets of functions used. 

One of the well-known representatives of these algorithms, with the best documentation and a large number of application examples, is pySINDy \cite{pysindy_art,pysindy_git}. The disadvantage of the library is that the sets of functions used cannot be set parametrically, and therefore in the reconstructed equations can be present only functions that differ from the functions from the a priori set by a multiplicative multiplier, parametric assignment of functions is impossible.

The FEDOT-EPDE framework \cite{epde_art,epde_git} does not have this drawback, it is also based on sparse regression, but it uses an evolutionary algorithm to select the necessary functions from a parametrically defined set, this allows to define parametrically classes of functions, which in turn allows to use the algorithm flexibly and to search for the necessary equations. However, when recovering partial derivative equations, the algorithm cannot recover mixed derivatives due to the way the numerical derivatives are calculated.

The SubTSBR \cite{SubTSBR_art,SubTSBR_git} algorithm, through the use of thresholded sparse Bayesian regression, can recover equations from data with high noise and outliers. The \cite{Robost_art,Robost_git} code uses low rank sequential (group) thresholded sparse Bayesian regression for the same task. The DySMHO library \cite{DySMHO_art,DySMHO_git} is based on a \textbf{moving horizon principle} for extracting the necessary functions from the base libraries, which allows increasing the size of libraries while maintaining the search time. The \cite{PSI_PDE_art,PSI_PDE_git} combines noise filtering by neural networks and search of governing equations by sparse regression.

\subsection{Methods based on neural networks }
A next class of algorithms reconstruct only certain types of equations and are weakly noise and emission tolerant, so the data must be well prepared. The best known, in terms of number of citations, is the PDE-Net-1 \cite{pde_net1_art,pde_net1_git} algorithm, which is based on the coupling of ultra-precise neural layers with differential operators. It recovers only parabolic equations and requires a considerable amount of data for good recovery.
The PDE-Net-2 \cite{pde_net2_art,pde_net2_git} code is a logical extension of the previous one, with the addition of a symbolic multilayer neural network to recover governing equations
The HPM algorithm \cite{HPM_art,HPM_git} utilises the coupling of neural networks with numerical Gaussian processes to recover the model. The NeuraDiff \cite{neuradiff_art,neuradiff_git} code uses two neural networks: one network is used to extract values of physical model variables from experimental data, and the other neural network finds the temporal evolution of the physical model. By comparing the results of the neural networks, the desired model is reconstructed. DeepXDE library \cite{deepxde_art,deepxde_git} is oriented on physics-informed learning and allows to reconstruct some classes of differential equations.
\subsection{Other methods}
There are also other methods for obtaining governing equations, e.g. \cite{symbol_art} uses symbolic regression based on the graph of calculations, \cite{PDE_IF_art, PDE_IF_git} first extracts the characteristics of the data and then makes equations based on them. The library \cite{Green_art,Green_git} allows to find Green's functions, which can be useful in certain cases

The libraries described above allow to recover governing equations if the full solution of these equations is known. However, as mentioned above, there are cases when the value observed in the experiment describes only a part of the solutions of the system or a combination of these solutions, and the remaining solutions are unknown. Such a situation occurs, quite often in complex systems. For example, brain activity is described by a system of differential equations. The parameters that are visible by measuring devices are electrophysiological data obtained by electroencephalography (EEG) or electrocorticography. Such data describe an incomplete combination of solutions of the model system. This leads to the problem of how to reconstruct the complete model having only a certain part of the solutions. Obviously, the problem is non-trivial and it is impossible to solve it without a priori knowledge due to the complexity of the system and ambiguity of possible results. In \ref{sec:ProbDef} the full problem formulation was described.

\section{Problem definition} \label{sec:ProbDef}

In the brain's intricate neural communication, electrical impulses and chemical transmitters play vital roles. Non-invasive yet cost-effective techniques for investigating brain function include electroencephalography (EEG), which captures electrical potential from the scalp, and electrocorticography (ECoG), involving electrodes implanted directly into the brain for more detailed recordings.

\begin{figure}[h!]
\centering
\includegraphics[width=10cm]{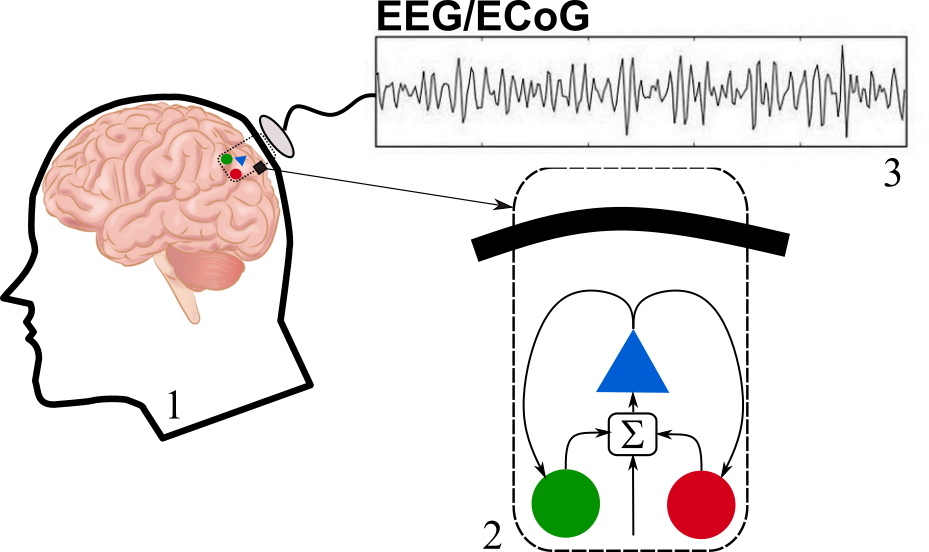}
\caption{Schematic representation of Jansen's \cite{Jansen1995} model. Currents in pyramidal neurons (blue pyramid), due to gain and attenuation circuits (populations of insertion neurons) generate the signal read on electroencephalograms.}\label{inint}
\end{figure}

The fundamental principle of EEG signals suggests that the captured electrical activity closely reflects the collective dynamics of pyramidal neurons, particularly their average firing rate over time. The classical resting-state model, exemplified by $\alpha$-waves, is based on a single-column model of pyramidal cells, as described by \cite{Jansen1995}. This model involves two interneuron-driven feedback loops: one boosts the signal within the pyramidal cells (green circle), while the other dampens it (red circle). Figure \ref{inint} offers a simplified depiction of this system, illustrating how the population of pyramidal cells (blue pyramids) is influenced by external signals from deeper brain regions, as well as the self-regulating dynamics through interneuron interactions that either amplify (green) or attenuate (red) the signal within the column.

Mathematically, such a model can be represented by a system of three equations:
\begin{equation}
\begin{cases}\label{eq:inin}
  (y_0)''+\gamma_0(y_0)'+\omega^2_0y_0=g_0(y_1-y_2)\\
  (y_1)''+\gamma_1(y_1)'+\omega^2_1y_1+f_1(t)=g_1(y_0)\\
  (y_2)''+\gamma_2(y_2)'+\omega^2_2y_2+f_2(t)=g_2(y_0)
\end{cases}\,.
\end{equation}
where $y_0$, $y_1$, $y_2$ are postsynaptic potentials after pyramidal, enhancing and attenuating interneurons, functions $g_i(y)$ are characterise the transformation of the average membrane potential of a population of neurons into the average density of impulses in the population of neurons. The signal (average pulse density) in pyramidal neurons is proportional to the difference $y_1 - y_2$, and hence the signal taken by electrodes located near the population is also proportional to $y_1 - y_2$.

On the basis of such a description, a problem arises which can be formulated in a general way as follows. There exists a process model described by a system of equations (\ref{eq:inin}), which are not completely known, in terms of parameters and/or functions. It is necessary to recover this system completely, knowing only the combination of its solutions in the form $y_1-y_2$.

The solution of such a problem will allow, on the one hand, to specify the models and parameters, and on the other hand, in the case of the model under consideration, to measure the signal external to this model, characterising the impact of deeper structures on the system. Also successes of such approach will give an opportunity to solve more complex problems, for example, to restore the model consisting of a larger number of equations \cite{big_models}.

\section{Our approch}
The main problem of the problem is that the number of visible (measurable) variables is less than the number of unknown equations. In general, it is impossible to solve such a problem or it is possible to obtain many solutions. Consequently, we assume that part of the equations or/and some basic characteristics are known a priori. This information can be obtained either from reference books and articles\cite{Jansen1995}, or by investigating solutions as in the \cite{PDE_IF_art}. For example, to understand the fundamental frequencies present in an equation one can study the Fourier image of the solution and construct and study spectrograms. 

Suppose that it is only possible to measure the value $y_1-y_2$ and hence its dependence on time is known. In the problem of building a model of the EEG signal, this value is obtained by measurement using microelectrodes. The general form of the model equations is also known:

\begin{equation}
    \begin{cases}
      (y_0)''+\gamma_0(y_0)'+\omega^2_0y_0=g_0(y_1-y_2)\\
      (y_1)''+\gamma_0(y_1)'+\omega^2_0y_1=g_1(y_0)\\
      (y_2)''+\gamma_2(y_2)'+\omega^2_2y_2+f_2(t)=g_2(y_0)\label{eq_model}
    \end{cases}
\end{equation}

For certainty, we consider that the first and second equations in the system (\ref{eq_model}) are known a priori, as well as all functions $g_i(y)$, the third equation is not known, we know that it may be an oscillatory equation with some source of general form. If the third equation is known, but the second one is unknown, then the sequence of steps of the algorithm is similar with the change of equation numbers.

We can say that the purpose of the algorithm is, on the one hand, to refine the model and introduce additional terms that may describe the specific behaviour of the EEG or electrocorticography signal, and, on the other hand, to describe the input signal to the model, which may be due to neural activity of the inner layers of the brain that do not have a direct influence on the measured signals.

The algorithm follows these steps:
\begin{enumerate}
    \item Given a known first equation of the system (\ref{eq_model}) and a time-dependent value $y_1-y_2$, the first equation is solved with different initial conditions. The initial conditions for $y_0$ are taken in the range $(min(y_1-y_2)-\epsilon,max(y_1-y_2)+\epsilon)$, for $y'_0$ in the range $(min(y'_1-y'_2 )-\epsilon', max(y'_1-y'_2)+\epsilon')$. The values $y_0$, $y_1$ and $y_2$ are assumed to be of the same order, $\epsilon$ and $\epsilon'$ are hyperparameters of the model. The result of the first step is the set of solutions of the first equation of the system (\ref{eq_model}) for different initial conditions -- $y_{0\; m}$.
    \item The second step is to solve the second equation of the system (\ref{eq_model}) for different initial conditions for $y_1$, which are taken similarly to the initial conditions from the last step, for all values of $y_{0\; m}$ obtained in the last step. The result of the second step is the set of solutions of the second equation of the system (\ref{eq_model}) for different initial conditions -- $y_{1\; m \; k}$, where the first index $m$ corresponds to the initial condition for $y_0$, and the second index $k$ corresponds to the initial condition for $y_1$.
    \item In the third step, the possible solutions to the third equation $y_{2\;m\;k} = - (y_1-y_2)-y_{1\;m\;k}$ are calculated, and then the third equation from the system is recovered using these solutions. There may be many different recovered equations, so an important step is to identify the equation that is needed.
    \item At the final step, the most appropriate equation for the system is selected from the set of reconstructed control equations on the basis of different criteria. During the experiments, the best criteria for selecting equations close to the required equations were the convergence criterion (the system converged faster to the required equations) and the boundaries of some parameters of the reconstructed equations.
\end{enumerate}

\section{Experiments}
\subsection{Data description}
\begin{figure}[!h]
\centering
\includegraphics[width=15cm]{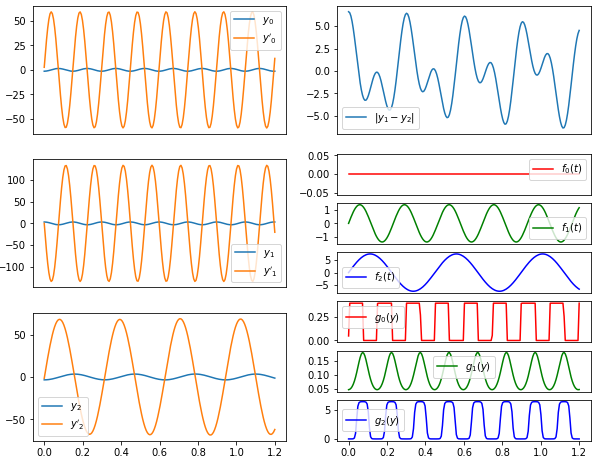}
\caption{Example of synthetic data. Left column of graphs -- hidden variables, upper right graph of the variable visible by the microelectrode, remaining graphs on the right -- graphs of functions $g_i(t)$ and $f_i(t)$}\label{fig:syn_examp}
\end{figure}

Two types of data were used in the experiments. The first type includes synthetic data. These are data obtained by solving a system describing a model of pyramidal cell population behaviour, from which a signal visible by microelectrodes can be obtained. For simplicity, attenuation was not considered in the model. The model equations were as follows:
\begin{equation}
    \begin{cases}
      (y_0)''+\omega^2_0\;y_0+f_0(t)=g_0(y_1-y_2)\\
      (y_1)''+\omega^2_0\;y_1+f_1(t)=g_1(y_0)\\
      (y_2)''+\omega^2_2\;y_2+f_2(t)=g_2(y_0)\label{eq_dataset}
    \end{cases}\; ,
\end{equation}
where $g_i(y) = \frac{C_i}{1+\exp(vi-\alpha_i\; y)}$\cite{Jansen1995}, the signal measured by the microelectrodes is $y_1(t)-y_1(t)$.
The synthetic data allow us to trace the reconstruction algorithm when the whole system is known, which will allow us to find the strengths and weaknesses of the algorithm, as well as to determine the selection criteria for the reconstructed governing equations.
Figure \ref{fig:syn_examp} shows a description of the synthetic dataset and the model that generates it. The left column of graphs shows the model variables $y_i$ that are hidden from measurement. The right graphs show the measured variable $(y_1-y_2)$ and the functions $f_i(t)$, $g_i(t)$. In the process of conducting the experiments, a lot of synthetic data corresponding to different models were used, the parameters of the used models are presented in tables \ref{table:no_force} and \ref{table:with_force} of Appendix A. The main principle of model selection was as follows: many models ($\sim 10^3..10^4$) with random parameters from the ranges close to real parameter values were generated, and then the signal ($y_1-y_2$), which resembles typical repeating patterns in real signals, was selected.

\begin{figure}[!h]
\centering
\includegraphics[width=15cm]{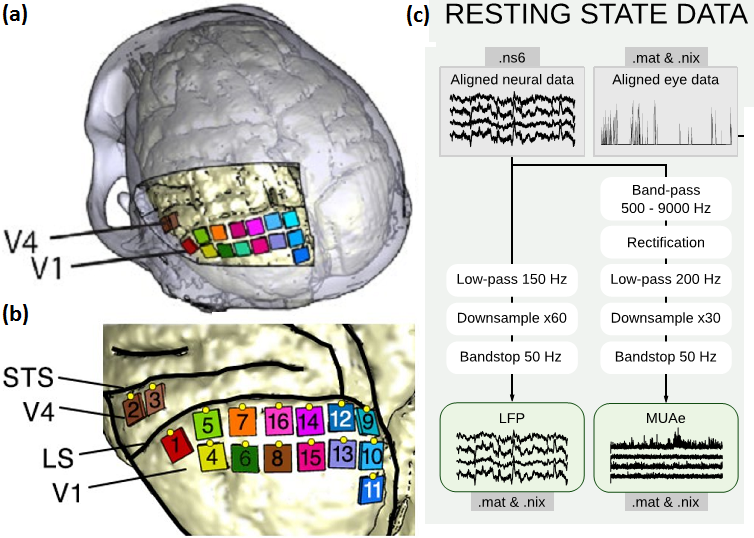}
\caption{Location of microelectrode arrays in visual cortex in areas V1 and V4. (a) -- general position of arrays, (b) -- exact position of microelectrode arrays, (c) -- signal post-processing (filtering, downsampling)) \cite{dataset}.}\label{fig:dataset}
\end{figure}

The second dataset was a set of electrophysiological recordings \cite{dataset} obtained from implanted electrodes in the primate visual cortex. The data describe the signal of the microelectrodes in the primate resting state. In the resting state, $\alpha$-waves are generated, a model of which is described in \cite{Jansen1995}. Data are obtained from 16 implanted arrays of The Utah microelectrodes of dimension $8\times8$. Figure \ref{fig:dataset} (a) and (b) show the location of the microelectrode arrays; inset (c) shows the postprocessing of the microelectrode signal, including filtering and downsampling).

\subsection{System of equations without impact}
The first series of experiments is the reconstruction of the system of equations describing the model when the parameters of two equations are known. For this case, the general system is as follows:
\begin{equation}
    \begin{cases}
      (y_0)''+\omega^2_0\;y_0=\frac{C_0}{1+\exp(v_0-\alpha_0(y_1-y_2))}\\
      (y_1)''+\omega^2_0\;y_1=\frac{C_1}{1+\exp(v_1-\alpha_1 \;y_0)}\\
      (y_2)''+\omega^2_2\;y_2=\frac{C_2}{1+\exp(v_2-\alpha_2\;y_0)}\label{eq:no_force}
    \end{cases}\; ,
\end{equation}
in this equation the values $\omega_0, \; C_i, \; v_i, \; \alpha_i $ are known, the unknown quantity is $\omega_2$.
\begin{figure}[h!]
\centering
\includegraphics[width=12cm]{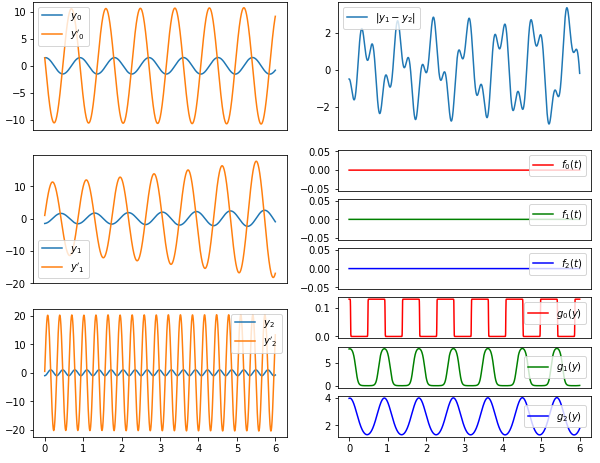}
\caption{ The model described by the system (\ref{eq:no_force}), where $\omega_0 = 7$, $\omega_2 = 20$, $C_0 = 0.13$, $C_1 = 9.0$, $C_2 = 9.0$, $\alpha_0 = 6$, $\alpha_1 = 3$, $\alpha_2 = 0.5$, $v_0 = 0.1$, $v_1 = 2.5$, $v_2 = 1.0$. Number of time steps is 2000, time interval [0, 6], vector of initial conditions $\textbf{y} = [1.5, 1.5, -1.5, 1.0, -1.0, 0.5]$. }\label{fig:ord_1_1}
\end{figure}

Figure \ref{fig:ord_1_1} shows the graphs describing the given behaviour of the model, the left shows the behaviour of the hidden variables of the model $y_i$, the upper right graph shows the visible signal, the other graphs on the right show the values of the functions $f_i(t)$ and $g_i(y)$ (see the system (\ref{eq:no_force}). This model is characterised by two frequencies $\omega_0 = 7$ and $\omega_2 = 20$. The other parameters of the model are presented in the caption to the figure.

\begin{figure}[h!]
\centering
\includegraphics[width=12cm]{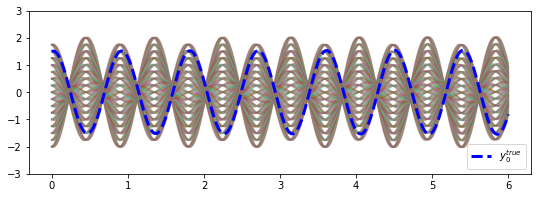}
\caption{Solutions $y_{0\;m}$ of the first equation of the model shown in Figure \ref{fig:ord_1_1}. The blue dashed line shows the true dependence of $y_0$.}\label{fig:ord_1_2}
\end{figure}

Let us consider the steps of the algorithm in sequence. First, using the known first equation of the system, the known function $g_0(y_1-y_2)$ and the known measured signal ($y_1-y_2$), we construct, using the explicit Runge-Kutta method of order 5, the solutions $y_{0\;m}$ of this equation for different initial conditions (we assume that the interval of initial conditions captures the desired initial conditions). Figure \ref{fig:ord_1_2} shows the solutions of the equation for different initial conditions, with the blue dashed line showing the desired behaviour of $y_0$.

\begin{figure}[h!]
\centering
\includegraphics[width=12cm]{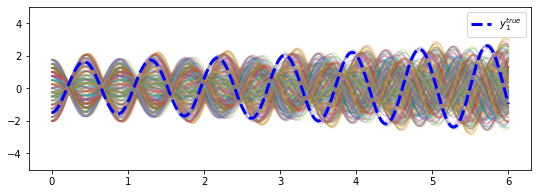}
\caption{Part of the solutions $y_{1\;m\;k}$ of the second equation of the model shown in Figure \ref{fig:ord_1_1}. The blue dashed line shows the true dependence of $y_1$.}\label{fig:ord_1_3}
\end{figure}

The next step is to solve the second equation for the set of initial conditions and for each obtained $y_{0\;m}$. These solutions $y_{1,\;m\;k}$ (index $m$ corresponds to the initial conditions for $y_0$ and index $k$ for $y_1$) are shown in Figure \ref{fig:ord_1_3}. As in the previous case, the initial conditions are chosen to capture the desired initial conditions.

The final step of the algorithm is to recover equations of the form:
\begin{equation}
      (y_2)''+\omega^2_2\;y_2=\frac{C_2}{1+\exp(v_2-\alpha_2\;y_0)}
\end{equation}

over all obtained possible solutions of this equation $y_{2\;m\;k}=y_{1\;m\;k} - (y_1-y_2)$, for the corresponding $y_{0\;m}$. 

\begin{figure}[h!]
\centering
\includegraphics[width=12cm]{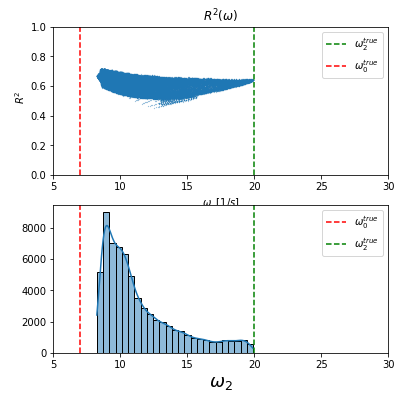}
\caption{The result of the algorithm for the model shown in Figure \ref{fig:ord_1_1}. The upper graph shows the distribution of the recovered equations on the $\omega_2-R^2$ plane. The lower graph shows the distribution of the recovered equations ok $\omega_2$. The red and green dashed lines show the true values of $\omega_0$ and $\omega_2$, respectively.}\label{fig:ord_1_4}
\end{figure}

Figure \ref{fig:ord_1_4} shows the result of the recovery of all differential equations. The upper plot shows the distribution of the recovered systems in the frequency $\omega_2$ -- coefficient of determination (for sparse regression) plane, the abscissa axis shows the frequency $\omega_2$ for the recovered equation, and the ordinate axis shows the coefficient of determination for the final value of the sparse regression at the resulting differential equation. The bottom figure shows a histogram of the distribution of the recovered equations by $\omega_2$ frequencies. The red and green dashed lines show the true values of $\omega_0$ and $\omega_2$, respectively. The figure shows, and subsequent experiments confirm this, that the obtained frequencies for the reduced equations $\omega_{2\;m\;k}$ (the set of frequencies $\omega_{2}$ for all reduced differential equations) lie between the frequencies $\omega_0$ and $\omega_2$ of the model, and if the true initial conditions are included in the initial conditions used in the algorithm, then the distribution $\omega_{2\;m\;k}$ touches the true frequency $\omega_2$.

\begin{figure}[h]
\begin{minipage}[h]{0.49\linewidth}
\center{\includegraphics[width=1.0\linewidth]{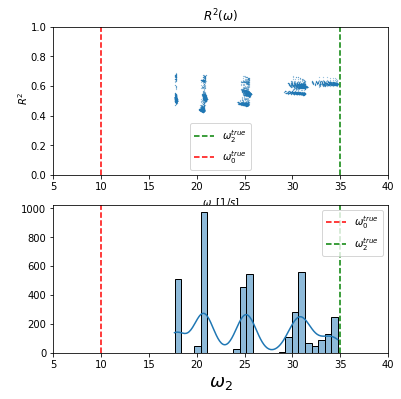} \\ (a)}
\end{minipage}
\hfill
\begin{minipage}[h]{0.49\linewidth}
\center{\includegraphics[width=1.0\linewidth]{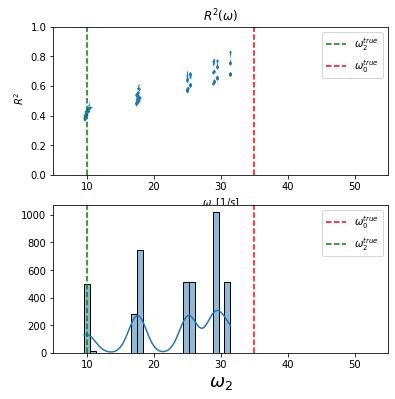} \\ (b)}
\end{minipage}
\caption{The results of equation reconstruction by models whose parameters are presented in the table \ref{table:no_force} row 1 and row 2, respectively. Upper graphs are distribution of the reduced differential equations on the plane $(\omega_2 ; R^2)$, lower graphs are histograms of the results by frequency $\omega_2$.}\label{fig:ord_23_4}
\end{figure}

Let us consider two more cases of restoring the system of equations describing the model. In the table \ref{table:no_force} the parameters of these models are presented under numbers 1 and 2, and the figure \ref{fig:ord_23_4} shows the results from left to right, respectively. In all cases, it can be seen that the frequency distribution of the recovered equations $\omega_{2\;m\;k}$ always touches the frequency $\omega_{2}$ if the initial conditions sought are included in those generated during the algorithm. This is confirmed by other experiments with different models that have been carried out, their parameters are presented in the tables in appendix A. This result probably has a strict mathematical justification, but let us take it as a heuristic rule that will help to work with the algorithm and identify the necessary equations.

Let us consider a series of experiments aimed at determining the influence of the accuracy of initial conditions when solving the first two equations in the system describing the model. Let us consider the system defined by the parameters of the 3rd row from the table \ref{table:no_force}. Let us take initial conditions close to the required ones and obtain solutions for them. Then we increase the interval of initial conditions for $y_0$, and leave it small for $y_1$, and get solutions for this case. Then we take a small spread in the initial conditions for $y_0$ and a large one for $y_1$. Finally, let us consider the algorithm's performance under the initial conditions for $y_0$ and $y_1$, which do not include the desired initial conditions. We will trace the influence of initial conditions on the final result, which will allow us to determine for which equations it is necessary to be more precise in setting the initial conditions during the operation of the algorithm.

\begin{figure}[!h]
\begin{minipage}[h]{0.47\linewidth}
\center{\includegraphics[width=1\linewidth]{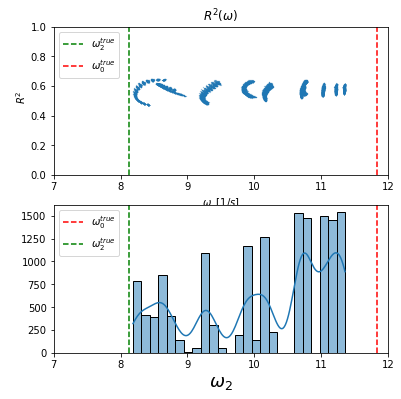}} (a) \\
\end{minipage}
\hfill
\begin{minipage}[h]{0.47\linewidth}
\center{\includegraphics[width=1\linewidth]{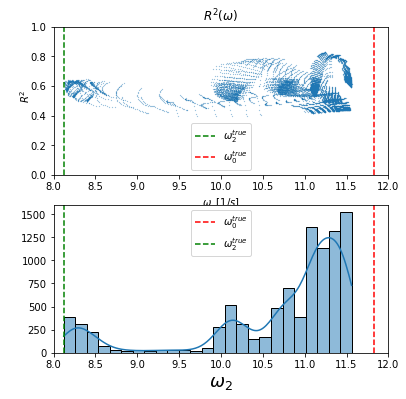}} \\(b)
\end{minipage}
\vfill
\begin{minipage}[h]{0.47\linewidth}

\center{\includegraphics[width=1\linewidth]{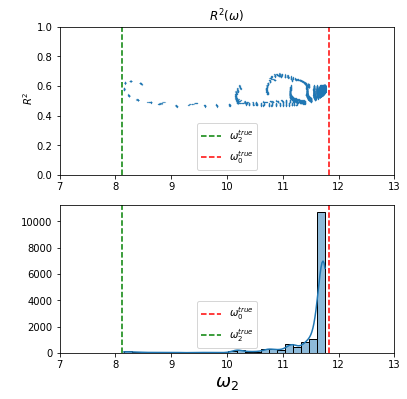}} (c) \\
\end{minipage}
\hfill
\begin{minipage}[h]{0.47\linewidth}
\center{\includegraphics[width=1\linewidth]{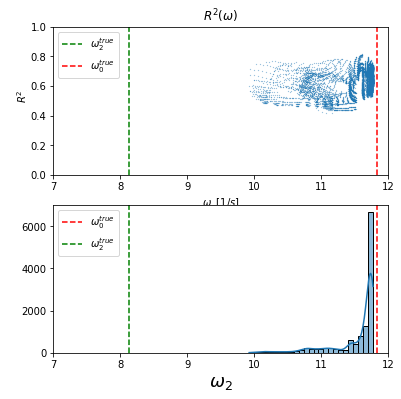}} (d) \\
\end{minipage}
\caption{Algorithm results, plane $(\omega_2;R^2)$ and histogram-$\omega_2$: (a) is narrow intervals of initial conditions for $y_0$ and $y_1$, the desired initial conditions are included, (b) is wide interval of initial conditions for $y_0$ narrow for $y_1$ the desired initial conditions are included, (c) is wide interval of initial conditions for $y_1$ narrow for $y_0$ the desired initial conditions are included, (d) is wide intervals of initial conditions for $y_0$ and $y_1$, the desired initial conditions are not included.}\label{fig:init_res}
\end{figure}

Figure \ref{fig:init_res} shows the results. From the results, we can conclude that if the desired initial conditions are included, the true value of $\omega_2$ is reached at the boundary of the distribution $\omega_{2\;m\;k}$. The effect of the imprecision of the initial conditions for $y_1$ is stronger than the imprecision of the conditions for $y_0$, hence the interval of the initial conditions for $y_1$ should be larger and more detailed than for $y_0$. If the desired initial conditions do not fall within the considered ones, then it is impossible to determine exactly $\omega_2$.

\subsection{System of equations with impact}

Let us consider the recovery of a system of differential equations describing models with influence. Such models may arise in situations where the model itself is underdetermined a priori with accuracy up to a function of time, or if the model is not closed and has an external connection, i.e. the system receives some external signal (influence). The system of equations describing such a model has the following form:
\begin{equation}
    \begin{cases}
      (y_0)''+\omega^2_0\;y_0+f_0(t)=\frac{Cg_0}{1+\exp(v_0-\alpha_0(y_1-y_2))}\\
      (y_1)''+\omega^2_0\;y_1+f_1(t)=\frac{Cg_1}{1+\exp(v_1-\alpha_1 \;y_0)}\\
      (y_2)''+\omega^2_2\;y_2+f_2(t)=\frac{Cg_2}{1+\exp(v_2-\alpha_2\;y_0)}\label{eq:with_force}
    \end{cases}\;.
\end{equation}

In the context of the model describing the population of pyramidal cells, the functions $f_i(t)$ can describe the effect on this population of other types of cells that are located in deeper regions of the brain. In the experiments we will assume that the values $\omega_0$, $\omega_1$, $Cg_i$, $v_i$, $\alpha_i$ are known, $f_0(t)=0$ and $f_1(t)=0$, the unknown function is $f_2(t)$, and it is known that it has the form $C_f\; \sin(\omega_f \; t)$, it is necessary to find its parameters. The choice of the appropriate function is due to the fact that the signal observed in the EEG has an oscillatory nature and probably the difficulty in restoring the governing equations will be in identifying and separating the various oscillatory terms. It is much easier to identify and describe trends (if they are present) in such a signal.
In the process of conducting experiments, it turned out that for true values of the functions the system recovery is faster and the convergence criterion reaches small values faster, so this criterion is used to guide the determination of the true governing equation. 

\begin{figure}[h!]
\centering
\includegraphics[width=12cm]{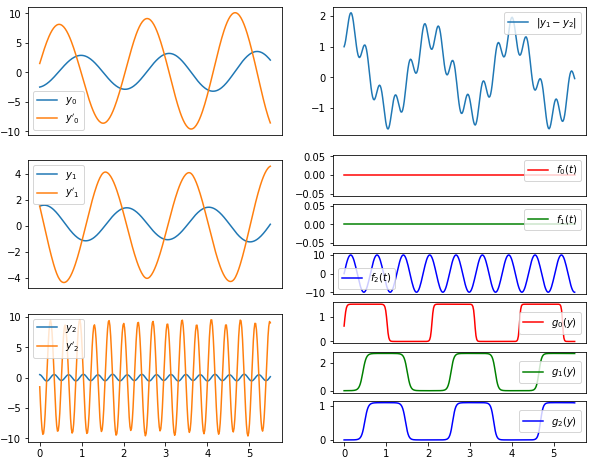}
\caption{The model described by the system (\ref{eq:with_force_exp}), wher $\omega_0 = 3.0$, $\omega_2 = 18.0$, $\omega_f = 10$, $Cg_0 = 1.5$, $Cg_1 = 2.7$, $Cg_2 = 1.1$, $C_f = 10.0$, $\alpha_0 = 2.1$, $\alpha_1 = 3.0$, $\alpha_2 = 3.1$, $v_0 = 0.35$, $v_1 = 0.13$, $v_2 = 1.31$. Number of time steps is 1000, time interval [0, 5.5], vector of initial conditions $\textbf{y} = [-2.5, 1.5, 1.5, 1.5, 0.5, -1.5]$. }\label{fig:force_md1}
\end{figure}

Let us consider a few experiments that will show the general approach to determining the governing equations. The system sought will be of the form:

\begin{equation}
    \begin{cases}
      (y_0)''+\omega^2_0\;y_0=\frac{Cg_0}{1+\exp(v_0-\alpha_0(y_1-y_2))}\\
      (y_1)''+\omega^2_0\;y_1=\frac{Cg_1}{1+\exp(v_1-\alpha_1 \;y_0)}\\
      (y_2)''+\omega^2_2\;y_2+C_f\; \sin(\omega_f \; t)=\frac{Cg_2}{1+\exp(v_2-\alpha_2\;y_0)}\label{eq:with_force_exp}
    \end{cases}\;.
\end{equation}
The last step of the algorithm will search for a differential equation of the form:
\begin{eqnarray}
 (y_2)''+\omega^2_2\;y_2+C_{f1}\; \sin(\omega_{f1} \; t)+C_{f2}\; \sin(\omega_{f2} \; t)+ \nonumber\\+C_{f3}\; \sin(\omega_{f3} \; t)=\frac{Cg_2}{1+\exp(v_2-\alpha_2\;y_0)}\;,\label{eq:3d_eq_with_force}
\end{eqnarray}

where one of the frequencies $\omega_{fi}$ coincides with the desired frequency $\omega_f$, the remaining two are different. The expected correct result of the reconstruction should be: $C_{fj} = C_{f}$ if $\omega_{fj}=\omega_f$ and $C_{fj} = 0$ if $\omega_{fj}\neq \omega_f$.
A description of the first model is shown in Figure \ref{fig:force_md1}.

\begin{figure}[!h]
\centering
\includegraphics[width=17cm]{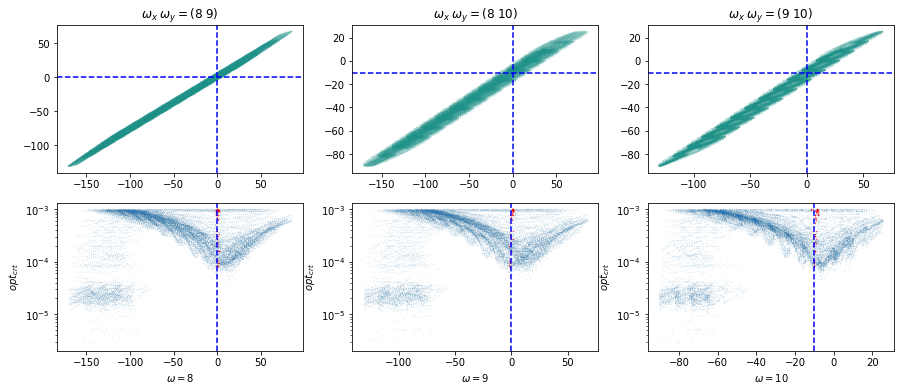}
\caption{Algorithm results for the model shown in Figure \ref{fig:force_md1}. Upper graphs represent the reduced equations on the $(C_{fi}\;,C_{fj})$ planes. Lower graphs representation of the reduced equations on the plane $(C_{fi}\;,opt_{crt})$, where $opt_{crt}$ is convergence criteria. The blue dashed lines represent the true values of $C_{fi}$.}\label{fig:res_md1}
\end{figure}

The result of the algorithm is shown in Figure \ref{fig:res_md1}, where the distribution of the results of the recovered equations on the plane $(C_{fi}\;,C_{fj})$ is presented on top. The bottom presents the distribution of the recovered equations on the planes $(C_{fi}\;,opt_{crt})$, where $opt_{crt}$ are the convergence criteria for recovering a particular equation of the system. The convergence criterion is related at the same number of iterations to the convergence rate. For this experiment, $\omega_{fi} = 8,\;9,\;10$. The blue dashed lines show the true values of the coefficients.

The parameters of the other models on which the experiments were performed are presented in the appendix in the table \ref{table:with_force}. The recovered parameters ($C_f$) of some models are presented in the appendix in Figures \ref{fig:res_md2} - \ref{fig:res_md5}.

As a result of the experiments we can conclude that despite the ambiguity in the initial conditions and the complexity of restoring the governing equations, the correct coefficients $C_{fi}$ in the differential equations are restored quite well, and the main criteria allowing to restore the desired system are the criteria related to convergence in the detection of equations

\subsection{Equation extraction from real data}

\begin{figure}[!h]
\centering
\includegraphics[width=17cm]{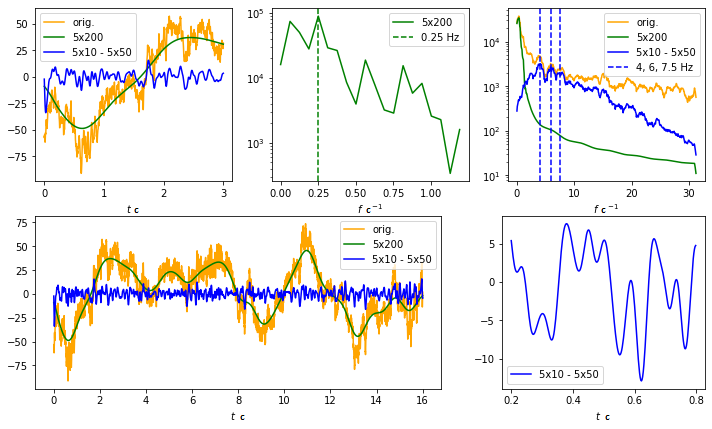}
\caption{The orange colour shows the electrophysiological [mV] signal obtained at resting state of the monkey by the implanted microelectrode of the Utah \cite{dataset} array with a sampling rate of 500 Hz. The low-frequency("5x200") and high-frequency("5x10-5x50") parts of the signal are shown in green and blue, respectively. All signals are presented in time and frequency representations.}\label{fig:real_data}
\end{figure}

Electrophysiological data acquired by microelectrode in a resting monkey, which was at rest with eyes closed, were used to reconstruct the equation describing the model. Articles were used to obtain a priori information and the microelectrode data were examined. Information from \cite{Jansen1995} was used to obtain information about the $g_i(y)$ functions, and Fourier transforms and spectrum studies were used to obtain the characteristic frequencies of the model.
Figure \ref{fig:real_data} shows the signal obtained from the microelectrode. Orange shows the true signal, green shows the low-frequency part of the signal (see figure caption for details), and blue shows the high-frequency part. The signals are shown in the time and frequency domains. The dotted lines show the highlighted frequencies. The low-frequency part of the signal ("5x200") is obtained sequentially by 5 running averages with a window of 200 elements. The high-frequency part of the signal is obtained as the difference of two parts. The reduced ("5x10") is the signal to which the running average with a window of 10 elements was applied 5 times in series, and the subtracted ("5x50") is the signal to which the running average with a window of 50 elements was applied 5 times in series

In the low-frequency signal it is 0.25 Hz, and in the high-frequency signal it is 4, 6, 7.5 and 17 Hz. Obviously, the overall signal is not a consequence of the model (\ref{eq_model}), which has only two frequencies, so let us consider the low-frequency and high-frequency parts separately. The low-frequency part is dominated by the frequency 0.25 Hz, which we will take as $\omega_0$. In the low-frequency spectrum reaches a maximum at frequencies 4 Hz - 7 Hz, but there is one more frequency in the spectrum, at which the maximum is reached on one side, and on the other side it corresponds to the frequency from the model \cite{Jansen1995} it is $\sim$ 17 Hz, we will take it as $\omega_0$ to restore the parameters of the model.

The low-frequency part of the signal corresponds to the model described by the system (\ref{eq:no_force}), for the frequency $\omega_0$ we take the frequency $2\pi\times 0.25$, shown in Figure \ref{fig:real_data} (green dashed line) and corresponding to the maximum of the spectrum of the low-frequency component.

\begin{figure}[h!]
\begin{minipage}[h]{0.49\linewidth}
\center{\includegraphics[width=1.0\linewidth]{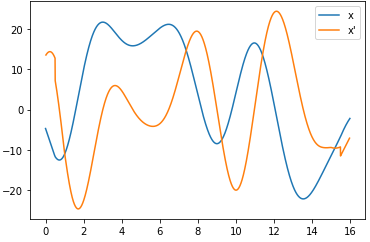} \\ (a)}
\end{minipage}
\hfill
\begin{minipage}[h]{0.49\linewidth}
\center{\includegraphics[width=1.0\linewidth]{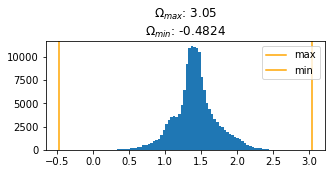} \\ (b)}
\end{minipage}
\caption{(a) is low-frequency part of the electrophysiological signal (b) is distribution of the $\omega_{2\;m\;k}$}\label{fig:ord_23_1}
\label{fig:real_low_comp}
\end{figure}

\begin{figure}[h!]
\begin{minipage}[h]{0.45\linewidth}
\center{\includegraphics[width=1.0\linewidth]{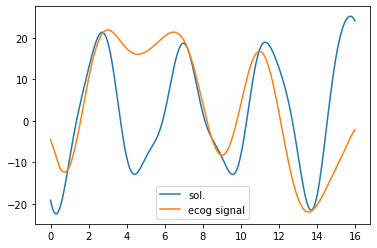} \\ (a)}
\end{minipage}
\hfill
\begin{minipage}[h]{0.45\linewidth}
\center{\includegraphics[width=1.0\linewidth]{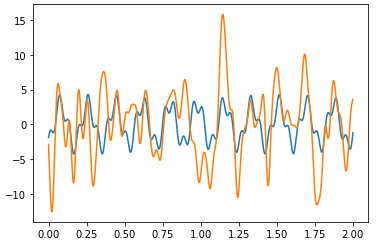} \\ (b)}
\end{minipage}
\caption{(a) shows a comparison of the solution of the system (\ref{eq:no_force}) at $\omega_0 = 1.5$ and $\omega_2 = 3.0$ (blue curve) with the low-frequency part of the electrophysiological signal (orange curve). (b) is a comparison of the solution of the system (\ref{eq:no_force}) at $\omega_0 = 100$ and $\omega_2 = 35$ (blue curve) with the high-frequency part of the electrophysiological signal (orange curve).}\label{fig:fig:real_result}
\label{fig:real_comp}
\end{figure}

Figure \ref{fig:real_low_comp}(a) shows the time-dependent low-frequency component $x$ of the signal, and \ref{fig:real_low_comp}(b) shows the result of the algorithm -- the distribution $\omega_{2\;m\;k}$. The boundary frequency is $\omega_2=3.05$. For frequencies $\omega_0 = 1.5$ and $\omega_2 = 3.0$, the solution of the system (\ref{eq:no_force}) was plotted, comparing it with the low-frequency component shown in Figure \ref{fig:real_comp}(a). The solution does not correspond exactly to the desired signal, but the overall similarity is clear, given the large uncertainty in the a priori data, the correspondence between the signals is very good, i.e. we can say that we have recovered the system of differential equations and it can serve as a relatively reliable model describing the desired signal.

\begin{figure}[!h]
\centering
\includegraphics[width=14cm]{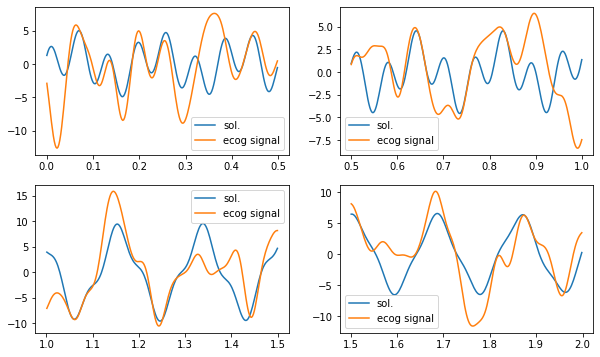}
\caption{Comparison of the solution of the system (\ref{eq:no_force}) at $\omega_0 = 100$ and $\omega_2 = 35$ (blue curve) with the high-frequency part of the electrophysiological signal (orange curve), for signal intervals.}\label{fig:cut_result}
\end{figure}

For the high-frequency part of the spectrum we proceed similarly, but $\omega_0 = 2\pi\times 17$. We get $\omega_2=35$. Figure \ref{fig:real_comp}(b) shows the solution for the system obtained by the algorithm for the high-frequency part of the signal. The obtained solution partially repeats the desired one, but the inaccuracy seems to be a consequence of the presence of several harmonics in the spectrum of the solution, which indicates a more complex model. If we assume that for the high-frequency component there are periodic influences changing the signal, then breaking the total high-frequency sought signal into 4 intervals and finding separate solutions for each, we can significantly improve the matches (see Fig.\ref{fig:cut_result}).

\section{Conclusion}
The result of this research is an innovative algorithm that successfully tackles the difficult task of reconstructing the system of equations governing EEG signal generation from data. The algorithm overcomes inherent complexity and ambiguity by integrating a priori knowledge and numerical analysis. Key criteria for equation identification were uncovered through its application on synthetic data, along with the significant impact of initial conditions on the algorithm's results. The algorithm demonstrates efficacy on real data by uncovering discernible patterns, suggesting its potential effectiveness in modeling brain electrophysiological activity. 

However, it is important to acknowledge that the accuracy of equation recovery depends heavily on the quality of initial conditions and the a priori information used. Future advancements in the algorithm could involve more complex brain models with a larger number of equations, as well as refining the approach for models with fewer a priori constraints.

\section*{Acknowledgments}
The author expresses gratitude to the scientific supervisor, Vsevolod Chernyshev, for his guidance, support, and useful advice. Additionally, Alexandra Razorenova provided valuable comments on the work.

\bibliographystyle{unsrt}  
\bibliography{Main}

\clearpage
\newpage
\appendix
\section*{Appendix A. Parameters of the models}
\addcontentsline{toc}{section}{Appendix A}

\begin{table}[h!]
\centering
\begin{tabular}{c | c c c c c} 
 \hline
 № & $\omega_0$ & $\omega_2$ & $C_i$ & $\alpha_i$ & $v_i$ \\ [0.5ex] 
 \hline
 1 & 10.0 & 35.0 &  [7.5, 6.0, 8.5] & [4.5, 3.3, 3.8] & [1.1, 3.0, 4.5] \\ 
 2 & 35.0 & 10.0 &  [7.5, 6.0, 8.5] & [4.5, 3.3, 3.8] & [1.1, 3.0, 4.5] \\ 
 3 & 11.8 & 8.1 &   [5.9, 4.7, 6.2] & [1.6, 5.0, 5.6] & [3.7, 4.0, 2.4] \\   
  & 7.0 & 20.0 &   [0.13, 9.0, 9.0] & [6.0, 3.0, 0.5] & [0.1, 2.5, 1.0] \\ 
  & 29.2 & 15.0 &   [1.5, 2.7, 1.1] & [2.1, 3.0, 3.1] & [0.35, 0.13, 1.3] \\ 
  & 10.0 & 17.0 &   [1.5, 2.7, 1.1] & [2.1, 3.0, 3.1] & [0.35, 0.13, 1.3] \\  
  & 17.0 & 10.0 &   [1.5, 2.7, 1.1] & [2.1, 3.0, 3.1] & [0.35, 0.13, 1.3] \\  
  & 15.0 & 27.0 &   [1.5, 2.7, 1.1] & [2.1, 3.0, 3.1] & [0.35, 0.13, 1.3] \\  
  & 10.9 & 23.6 &   [3.8, 7.1, 5.2] & [2.8, 2.4, 7.7] & [0.35, 0.13, 2.3]] \\  
  & 7.7 & 20.2 &   [0.13, 8.1, 9.6] & [6.0, 3.0, 0.5] & [0.09, 2.62, 1.0] \\    
  & 6.0 & 21.0 &   [1.5, 2.7, 1.1] & [2.1, 3.0, 3.1] & [0.35, 0.13, 1.3] \\  [1ex] 
  
 \hline
\end{tabular}
\caption{Parameters of systems of the form (\ref{eq:no_force}) used in the paper.}
\label{table:no_force}
\end{table}

\begin{table}[h!]
\centering
\begin{tabular}{c | c c c c c c c} 
 \hline
 № & $\omega_0$ & $\omega_2$ & $\omega_f$& $C_f$ & $Cg_i$ & $\alpha_i$ & $v_i$ \\ [0.5ex] 
 \hline
 1 & 3.0 & 18.0 & 10.0 & 10.0 &  [1.5, 2.7, 1.1] & [2.1, 3.0, 3.1] & [0.35, 0.13, 1.31] \\ 
 2 & 14.0 & 8.0 & 4.0 & 10.0 &  [1.5, 2.7, 1.1] & [2.1, 3.0, 3.1] & [0.35, 0.13, 1.31] \\  
 3 & 3.0 & 28.0 & 8.0 & 10.0 &  [1.5, 2.7, 1.1] & [2.1, 3.0, 3.1] & [0.35, 0.13, 1.31] \\ 
 4 & 8.0 & 14.0 & 3.0 & 10.0 &  [1.5, 2.7, 1.1] & [2.1, 3.0, 3.1] & [0.35, 0.13, 1.31] \\ 
  & 3.0 & 17.0 & 15.0 & 10.0 &  [1.5, 2.7, 1.1] & [2.1, 3.0, 3.1] & [0.35, 0.13, 1.31] \\ 
  & 3.0 & 25.0 & 8.0 & 10.0 &  [1.5, 2.7, 1.1] & [2.1, 3.0, 3.1] & [0.35, 0.13, 1.31] \\ 
  & 8.0 & 14.0 & 3.0 & 4.0 &  [1.5, 2.7, 1.1] & [2.1, 3.0, 3.1] & [0.35, 0.13, 1.31] \\  [1ex] 
 \hline
\end{tabular}
\caption{Parameters of systems of the form (\ref{eq:with_force_exp}) used in the paper.}
\label{table:with_force}
\end{table}

\newpage
\section*{Appendix B. Reconstructed models}
\addcontentsline{toc}{section}{Appendix B}


\begin{figure}[!h]
\centering
\includegraphics[width=14cm]{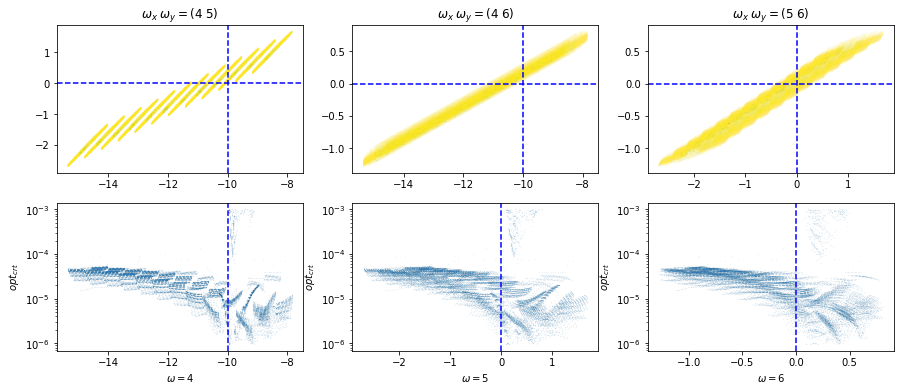}
\caption{Model extraction results -- table  \ref{table:with_force} row 2.}\label{fig:res_md2}
\end{figure}
\vspace{-0.5cm}

\begin{figure}[!h]
\centering
\includegraphics[width=14cm]{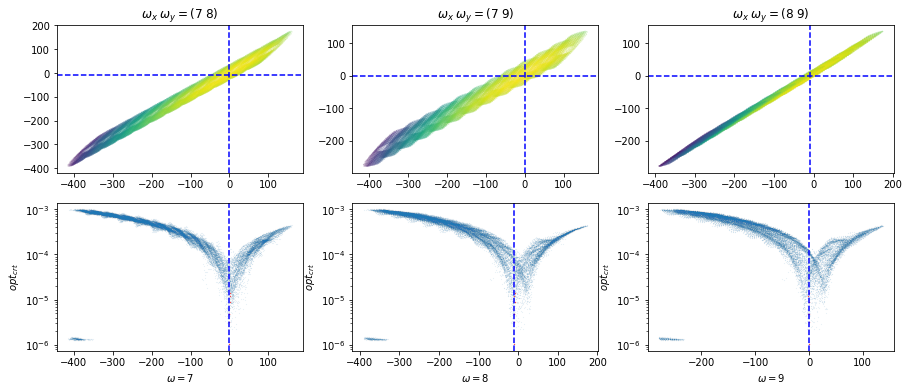}
\caption{Model extraction results -- table \ref{table:with_force} row 3.}\label{fig:res_md3}
\end{figure}
\vspace{-0.5cm}


\begin{figure}[!h]
\centering
\includegraphics[width=14cm]{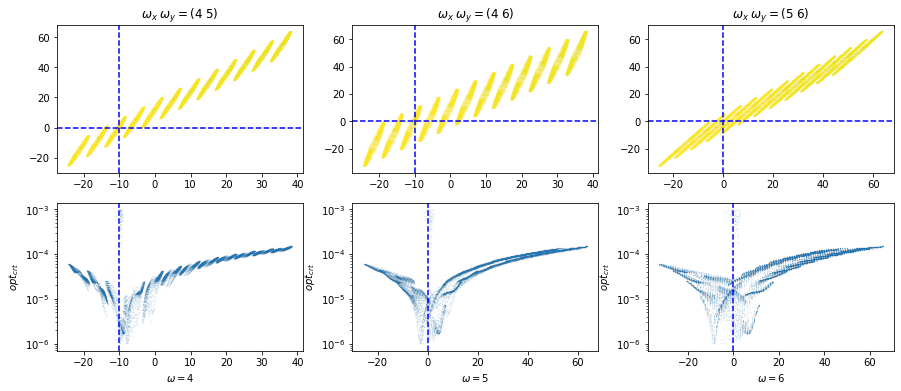}
\caption{Model extraction results -- table  \ref{table:with_force} row 4.}\label{fig:res_md5}
\end{figure}
\vspace{0.5cm}

\end{document}